\documentclass[aps,pre,twocolumn,superscriptaddress,floatfix,amsmath]{revtex4}
\usepackage{graphicx}
\usepackage{amsmath,amssymb}
\begin{document}
\title{Non-conservative forces and effective temperatures in active polymers}
\author{Davide Loi}
\affiliation{
European Synchrotron Radiation Facility, BP 220, F-38043 Grenoble 
France}
\author{Stefano Mossa}
\email[]{stefano.mossa@cea.fr}
\affiliation{UMR 5819 (UJF, CNRS, CEA) CEA, INAC, SPrAM,
17 Rue des Martyrs, 38054 Grenoble Cedex 9, France}
\author{Leticia F. Cugliandolo}
\email[]{leticia@lpthe.jussieu.fr}
\affiliation{Universit\'e Pierre et Marie Curie -- Paris VI, LPTHE UMR 7589,
4 Place Jussieu,  75252 Paris Cedex 05, France}
\date{\today}
\begin{abstract}
We follow the dynamics of an ensemble of interacting self-propelled
semi-flexible polymers in contact with a thermal bath. We characterize
structure and dynamics of the passive system and as a function of the
motor activity. We find that the fluctuation-dissipation relation
allows for the definition of an effective temperature that is
compatible with the results obtained by using a tracer particle as a
thermometer.  The effective temperature takes a higher value than the
temperature of the bath when the effect of the motors is not
correlated with the structural rearrangements they induce. Our data
are compatible with a dependence upon the square of the motor strength
(normalized by the average internal force) and they suggest an
intriguing linear dependence on the tracer diffusion constant times
the density of the embedding matrix.  We show how to use this concept
to rationalize experimental results and suggest possible innovative
research directions.
\end{abstract}
\maketitle
\section{Introduction}
\label{sect:Introduction}
The constituents of active matter, be them particles, lines or other,
absorb energy from their environment or internal fuel tanks and
use it to carry out motion. In this new type of soft condensed matter 
energy is partially transformed into mechanical work and partially dissipated 
in the form of heat~\cite{active-matter-reviews}. The units interact 
directly or through disturbances propagated in the medium.  

In systems of biological interest, conservative forces (and thermal fluctuations) 
are complemented by non-conservative forces. Realizations of active matter in biology 
are thus manifold and exist at different scales. Some of them are: 
bacterial suspensions~\cite{bacterial,bacterial-th,Vicsek,Czirok}, the cytoskeleton in living 
cells~\cite{howard01,alberts94,cyto,mackintosh10,deng06}, 
or even swarms of different animals~\cite{swarms}.  

Clearly enough, active matter is far from equilibrium and typically kept in a 
non-equilibrium steady state. The difference between active matter and other driven 
systems, such as sheared fluids, vibrated granular matter and driven vortex lattices 
is that the energy input is located on internal units ({\it e.g.} motors) and therefore 
homogeneously distributed in the sample. In the other driven systems just mentioned, 
the energy input occurs on the boundaries of the sample. Moreover, the effect of the 
motors can be dictated by the state of the particle and/or its immediate neighborhood 
and it is not necessarily fixed by an external field.

The dynamics of active matter presents a number of interesting features 
that are worth mentioning  here. Active matter displays 
out of equilibrium phase transitions that may be absent in their passive counterparts.
The dynamic states display large scale spatio-temporal dynamical patterns
and depend upon the energy flux and the interactions between their constituents. 
Active matter often exhibits unusual mechanical properties, very large responses to small
perturbations, and very large fluctuations -- not consistent with the central limit theorem. 

Much theoretical effort has been recently devoted to the description
of different aspects of these systems.  The investigation of the
self-organization of living microorganisms is summarized
in~\cite{BenJacob}.  The identification and analysis of states with
spatial structure, such as bundles, vortexes and asters was developed
in, {\it e.g.},~\cite{Nedelec97,Surrey,Kruse}.  The study of the
rheological properties of active particle suspensions was carried out
in~\cite{Hatwalne04,marchetti,tailleur,lau09} with the aim of grasping
which are the mechanical consequences of biological activity.

Surprisingly enough, molecular dynamics computer simulations, that have proven to be so 
helpful to probe and understand the dynamics of complex passive
systems, have not been much employed in the field of active matter yet. 
(For a recent review of the most used computational approaches to cell mechanics 
see Ref.~\cite{vaziri08}.) Studies presented so far have been mainly analytical, 
based on refined calculations on very stylized 
models~\cite{bacterial-th,Vicsek,swarms,BenJacob,Nedelec97,Surrey,Kruse,Hatwalne04,marchetti,tailleur}, 
computer simulations of continuum models, {\em e.g.}, 
Lattice Boltzmann equations~\cite{Ignacio}, and numerical studies of relatively simple lattice
models as the ones pioneered by Vicsek and collaborators~\cite{Vicsek,Czirok,Chate,Gonnella}. 

Quite independently, passive systems with complex out of equilibrium dynamics 
have been the focus of much attention too. 
These are systems with competing interactions between their constituents
-- and no external energy input -- 
that in some conditions -- such as sufficiently low temperature or high density -- 
cannot equilibrate with their 
environment. A regime (or phase) with glassy properties 
then occurs. In the last two  decades or so,  a rather complete mean-field theory that describes, 
up to a certain extent, the super-cooled liquid and deep glassy regime, 
developed~\cite{mct,cugliandolo02,cavagna,Binder-Kob,Barrat-Baschnagel-Lyulin}. 
Keywords are ``mode-coupling theory`` and the ``random first order phase transition 
scenario''. This has been complemented with extensive molecular dynamic simulations  of microscopic 
models with interactions between their constituents that are as realistic as 
possible~\cite{Binder-Kob,Barrat-Baschnagel-Lyulin}. Simulations have shown agreement with 
certain aspects of the theoretical approach, especially the ones that will be of interest in this paper. 

An important outcome of such theoretical modeling of glassy systems, later confirmed
with numerical simulations, is the generation of an effective temperature, $T_{eff}$, in super-cooled
liquids driven out of equilibrium and glasses in relaxation~\cite{cugliandolo97,cugliandolo-review}. 
The concept of effective temperatures in out of equilibrium 
systems, as arising from deviations from the fluctuation-dissipation 
theorem (FDT)~\cite{marconi08}, appeared in studies of dynamical systems~\cite{Hohenberg} and 
it was later extended and developed in the field of glassy systems~\cite{cugliandolo97}.
It was here fully appreciated that $T_{eff}$ should be understood as a 
time-scales dependent parameter and that it takes a thermodynamic 
meaning in systems with slow dynamics only~\cite{cugliandolo97,ilg06}. It was later proven that in systems 
with sufficiently slow dynamics -- reaching a regime of small entropy production --
the same values of the effective temperature are obtained with a variety of 
other measurements. 
Experimental measurements of $T_{\sc eff}$ in different glassy systems, including 
glycerol~\cite{Teff-glycerol}, spin-glasses~\cite{Teff-spinglass},
polymer samples~\cite{Teff-exp-pol}, granular matter~\cite{Teff-granular,Makse-granular}, 
laponite~\cite{Teff-laponite} and other colloidal suspensions~\cite{Teff-exp-coll}
have been performed. Very recently, a study of $T_{eff}$ in an active colloidal
suspension of Janus particles under a gravity field appeared in~\cite{palacci10} and 
we shall comment on this work later in the context of our results. On the 
numerical side, the work in~\cite{joly10} is especially relevant to our study and we 
shall also comment on it below. The validity of FDT  in the Vicsek model 
was investigated in~\cite{Czirok,Chate}. In the latter paper a dynamic 
measurement, similar to one of the calculations we present below, was 
performed. 

In~\cite{loi08} we took a first step in the direction of analyzing the
structure and dynamics of active matter with molecular dynamics
numerical methods.  We studied the dynamics of an ensemble of
self-propelled particles in contact with an equilibrated thermal bath,
a reasonable model for real but simple active matter, such as
bacterial colonies. In particular, we analyzed the generation of an
effective temperature and some of its properties. In this article we
go beyond this study to get closer to more complex cases. We introduce
a model for biological structures made of filamentous semi-flexible
polymers~\cite{liverpool06}. We analyze its structure and dynamics and
we study the effective temperature by using a variety of independent
measurements.  In particular, we show how to measure the effective
temperature by using techniques that have already been exploited in
the study of the mechanical properties of real biological systems.
The idea is to mimic real experiments which probe mechanical
properties of the embedding matrix by following the dynamics of tracer
particles (see, among many others, Refs.~\cite{legoff02,bursac05,wong04}), both
free and driven by external fields, {\it e.g.}, optical tweezers.  We
demonstrate that these techniques can be useful to obtain a direct
characterization of the out-of-equilibrium state, and allow for the
determination of an effective temperature.  A short summary of some of
the results included in the present work can be found in
Ref.~\cite{loi11}.

The paper is organized as follows. 
In Section~\ref{sect:Methods} we give details of the model and the molecular dynamics 
computer simulations performed.
Section~\ref{sect:Passive} presents a study of the molecular model in its 
passive limit in which we determine the optimal parameters, in particular
the polymer length, to be used in the simulations of the active case.
In Section~\ref{sect:Results} we describe our results on the structure and dynamics
of the molecular active matter model. 
Section~\ref{sect:temperatures} contains a detailed analysis of the effective temperature
measured by different methods and a comparison with other studies in the literature.
Our entire set of data is discussed and put in more perspective in  
Section~\ref{sect:Conclusions}, where we also give some conclusions 
and a few hints for future work.
\section{Model and methods}
\label{sect:Methods}
Many complex biological systems are constituted of elongated fibrillar structures 
(filaments)~\cite{liverpool06}. 
Although one could expect to use the well-established polymer theory to describe such systems, 
this theory focuses on the limits of completely flexible or totally rigid polymers~\cite{degennes79,doi87}
and most biologically relevant polymers or filaments are in neither of these classes. They rather 
belong in a third intermediate category: the one of semi-flexible filaments with persistence 
and contour lengths~\cite{footnote1} of comparable magnitude. The contour length of these filaments 
is too small for them to form loops and knots, yet they are sufficiently flexible to have significant 
thermal bending.  These problems are much harder to deal with analytically~\cite{morse}
and, surprisingly enough, relatively little numerical work has been done so far.  
In this paper we start filling this gap by 
presenting molecular dynamics simulations of a simple but sufficiently realistic model mimicking the
behavior of semi-flexible (or semi-rigid) polymers in interaction.
\subsection{Semi-flexible polymer model}
We consider a model of a (passive) semi-flexible polymer in which the components have the shape of
filaments.  Each filament is a linear chain of beads.  The optimal values of the parameters have 
been fixed by preliminary calculations, detailed in~\cite{miura01}. We briefly motivate them here.

Polymer chains are coarse-grained and each segment is formed by a bead. Each bead is in 
contact with a thermal environment that is described by a random noise and a viscous 
drag. The $i$th monomer pertaining to chain labeled $a$ has coordinates ${\bf r}_{ia}$ and 
velocity ${\bf v}_{ia}$. The latter evolve following the Langevin equation:
\begin{equation}
m_i\dot{\bf v}_{ia}=-\xi m_{ia} {\bf v}_{ia}+ {\bf f}_{ia}^s+ 
{\bf f}^M_{ia}+\mbox{\boldmath$\eta$}_{ia}.
\label{eq:langevin}
\end{equation}
$\mbox{\boldmath{$\eta$}}_{ia}$ is a Gaussian white noise representing
thermal agitation with zero mean and variance 
$\langle \eta^\mu_{ia}(t) \eta^\nu_{jb}(t') \rangle = 2 \xi m_{ia} T
\delta(t-t') \delta^{\mu\nu} \delta_{ij} \delta_{ab}$, where $\mu,\nu=1,\dots,d$ 
and $d$ is the dimension of space ($d=3$ in what follows). The temperature of the bath is
$T$ and we set $k_B=1$ henceforth. The term $-\xi m_{ia} {\bf v}_{ia}$ is the
frictional force. We take into account the over-damped character of the
dynamics by considering a large value of the friction coefficient. 
The numerical integration of Eq.~(\ref{eq:langevin}) generates a time
evolution that is  consistent with the canonic ensemble (NVT) in thermodynamic equilibrium. 
Note that we completely disregard the effect of hydrodynamic interactions due to internal and/or
intermolecular degrees of freedom.
We consider identical beads, with masses $m_{ia}=m$ and diameter $r_{ia}=r_0$
for all $i$ and $a$. 

The deterministic mechanical conservative force on monomer $i$ pertaining to
polymer $a$ due to all other components is ${\bf f}^s_{ia}\equiv
\sum_{b=1}^{N_p} \sum_{j=1}^{N_m} {\bf f}^s_{ijab} = -\sum_{b=1}^{N_p}
\sum_{j=1}^{N_m} \mbox{\boldmath$\nabla$}_{ia} U_{tot}(r_{ij}^{ab})$, where
$r_{ij}^{ab}$ is the distance between monomer $i$ in polymer $a$ and
monomer $j$ in polymer $b$.  $N_m$ and $N_p$ are the total 
number of monomers per chain and the total number of polymers respectively.
All polymers have the same length $N_m$. We consider the interaction potential
\begin{equation}
\label{eq:pot}
U_{tot}(r)= U_{intra}(r)+ U_{inter}(r),
\end{equation}
in which the two terms are the intramolecular ($a\neq b$) and intermolecular ($a=b$) contributions, 
respectively. The exact form of the two contributions must be chosen in order to expand the region 
of stability of the liquid state, avoiding crystallization, and to tune the degree of flexibility 
of the polymer chains, thus inhibiting complete folding of the chains.

For the intermolecular interactions we choose
\begin{widetext}
\begin{eqnarray}
\label{eq:pot-inter}
U_{inter}(r_{ij})=\left\lbrace
4\epsilon 
\left[\left(\frac{\sigma}{r_{ij}}\right)^{12}-\left(\frac{\sigma}{r_{ij}}\right)^{6}\right]
 -U(r_c)\right\rbrace\theta(r_c-r),
\end{eqnarray}
\end{widetext}
{\it i.e.}, the usual Lennard-Jones interaction potential cut and
shifted to zero at $r_c$ by adding the constant value $-U(r_c)$. In
addition, we consider $r_c=2^{1/6}\sigma$, corresponding to the
position of the minimum of the pure Lennard-Jones potential, and
obtain a purely repulsive potential ($U(r_c)=-\epsilon$ in this
case). The Heaviside function $\theta$ assures that the potential is
zero for $r>r_c$.  This particular choice is done to obtain a polymer
density in the melting state lower than the one achieved by
considering attractive interactions, thus avoiding crystallization at
low temperatures~\cite{miura01}. $\epsilon$ has energy units and
controls the strength of the interaction. $\sigma$ has length units
and fixes the typical length scale of the system.

The intramolecular potential reads:
\begin{widetext}
\begin{eqnarray}
\label{eq:pot-intra}
U_{intra}(r_{ij}) =
\begin{cases}
k(r_{ij}-r_0)^2, & \text{if $i, j$ n.n.}\\
\left\lbrace 4\epsilon \left[\left(\frac{S}{r_{ij}}\right)^{12}-\left(\frac{S}{r_{ij}}\right)^{6}\right]-U(r_c)
\right\rbrace\theta(r_c-r), 
& \text{if $i, j$ n.n.n.}
\end{cases}
\end{eqnarray}
\end{widetext}
Connectivity of the chain is assured by harmonic springs of elastic constant $k$ and 
equilibrium length $r_0$, acting between nearest neighbor monomers. Chain rigidity is controlled 
by a repulsive Lennard Jones potential between next nearest neighbors, with $r_c=2^{1/6}S$ 
and again $U(r_c)=-\epsilon$. Finally, monomers separated by more than two bonds interact via 
Eq.~(\ref{eq:pot-inter}). The actual choice for the parameters $k$, $r_0$ and $S$ allows one to 
control the flexibility of the chains, ranging from semi-rigid to flexible chains~\cite{miura01}.  
In particular, it is evident that chain rigidity can be induced by a mismatch of the zero point energy 
position of the two terms in Eq.~(\ref{eq:pot-intra}) taken separately. 
Therefore, the relevant parameter is the ratio $r_0/S$. Miura {\em et al.}~\cite{miura01} found that the 
average cosine of the angle between two successive bond vectors along the chain equals $0.92$, 
{\it i.e.}, the chain is semi-rigid, for $S\simeq 2.5 r_0$. 
\subsection{The molecular motors}
A preliminary remark about the implementation of realistic molecular motors in a 
coarse-grained computer simulation study is obvious: no simple technique is able 
to take into account chemical activity in a molecular dynamics calculation. 
Having this in mind, particular care and physical insight has been put in the choice of a 
reasonable model for motor action on a particular monomer, say the $i$th in the $a$th chain, 
that we call ${\bf f}^M_{ia}$. 
Only a fixed fraction of polymers are provided with motors. This choice is mainly
due to the requirements that motor activity: i) should be homogeneously distributed 
in the sample; ii) should have a gentle but still detectable out-of-equilibrium
effect on the behavior of the system; iii) should stay in the linear response regime.

\begin{figure}[h]
\vspace{0.15cm}
\includegraphics[width=0.4\textwidth]{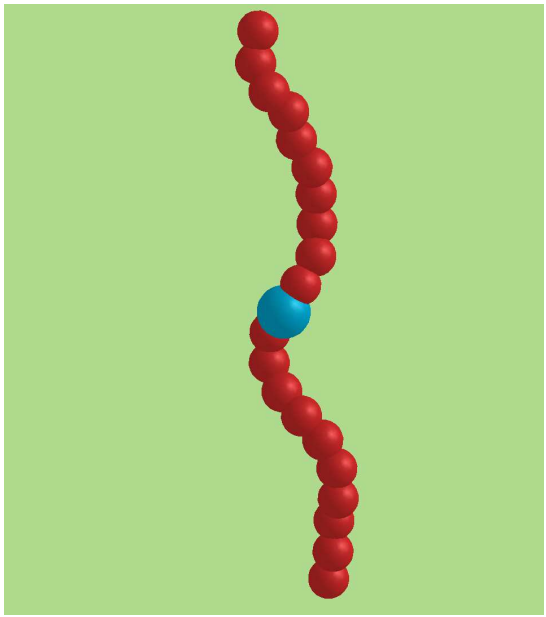}
\caption{
The semi-flexible polymer model considered in this paper. It is formed by 21 monomers 
among which 20 are non-motorized (red spheres). The entire chemical motor activity is
localized on the central monomer (blue sphere). Although all monomers have the same size 
we draw a larger central monomer in the sketch to emphasize its particular role.}
\label{fig:polymer} 
\end{figure}

Following symmetry considerations, we chose to localize the non-conservative polymer activity 
at the core of the filament, {\em i.e.}, we motorized the very central monomer in the polymer 
only. Note that, in general, this position does not coincide with the center of mass of the 
filament and, most importantly, it is fixed. This situation is clearly at variance with the case of
real systems, where molecular motors are able to slide along the filaments~\cite{howard01}. 
Nevertheless, we do not consider this a serious limitation to our approach, since we 
follow the dynamical behavior
of the system on limited time scales over which motor activity can be indeed taken as 
localized. Also, we average our results over a substantial number of different instances of 
motor realizations, mimicking in an indirect way motor motion relative to the filaments. 
A typical isolated flexible polymer configuration studied is shown in Fig.~\ref{fig:polymer}. 

Similar to the case of self-propelled spherical particles~\cite{loi08}, we mimic the motor 
effect with a time series of random isotropic kicks, generated by a suitable stochastic process 
controlled by three parameters: i) the number of motorized polymers; ii) the intensity and 
direction of the motors forces; and iii) the activation time scale. During $\tau$ steps of 
the molecular dynamics trajectory, independent forces are applied to a fraction of randomly 
chosen motorized central monomers. The strength of the force exerted,
$f^M=|{\bf f}^M_{ia}|$ is the same for all motors and is chosen to be a fraction of the mean conservative 
mechanical force acting on the equivalent passive system, $\overline F=(N_pN_m)^{-1} \sum_{a=1}^{N_p}
\sum_{i=1}^{N_m} |{\bf f}^s_{ia}|$.  We quantify the motor activity with the control parameter 
$f=f^M/\overline F$. The direction of the motor force is chosen at random isotropically, to avoid preferential
flows in the system.  The subset of propelled monomers and the directions of the applied forces change at
each power stroke of duration $\tau$. In short, we focus on {\em adamant}~\cite{shen04} motors, {\it i.e.} 
their action is completely independent of the structural rearrangements they induce.
\subsection{Simulation details}
In a nutshell, given a model for the interaction forces between the different coarse-grained units
forming a physical system, the technique of molecular dynamics simulation~\cite{allen89}
provide effective methods to numerically integrate the  equation of motion, conforming to 
different statistical ensembles and macroscopic external conditions. It has been utilized 
for some decades now, to explore structural and dynamical properties of complex systems 
and it has been shown to perform quite successfully also in the case of polymer 
systems~\cite{Barrat-Baschnagel-Lyulin,paul04,paul98} of interest here. 

In what follows we express all quantities in dimensionless reduced units: length, energy and 
time-units are $\sigma$, $\epsilon$ and $(m\sigma^2/\epsilon)^{\frac{1}{2}}$, respectively~\cite{allen89}.
All the monomers have the same mass $m=1$. In order to give an idea of typical realistic orders of 
magnitude, we can associate these dimensionless quantities to reasonable values for the experimentally 
measured counterparts~\cite{klimov99}. We consider as a typical unit of energy, $2 k_B T$. 
The typical lengths are measured in units of $0.4$~nm, which gives a typical order of magnitude for 
forces of $20$~pN at ambient temperature $T=300$~K.

Concretely, we integrated Eqs.~(\ref{eq:langevin}) numerically using Ermak's algorithm~\cite{ermak78,allen89},
with an integration time step $\delta t=0.001$ and  a friction coefficient $\xi=10$. This technique produces 
realistic configurations of the system, characterized by probability distributions conforming 
to the canonic NVT ensemble. We calculated statistical averages over several independent 
instances of the dynamics and we indicate them with angular brackets, $\langle \dots \rangle$.
We used a cubic simulation box of total volume $V$ with periodic boundary conditions.
The parameters in the interaction potentials are: $k=9000$ (the elastic constant), 
$r_0=0.4$ (the equilibrium spring length) and $S=1$. We simulated $N_p$ chains formed by $N_m$ monomers each. 
The total number of interaction sites is $N=N_p\times N_m$ and the number density 
$\hat\rho$ is given by the ratio $N/V$.
The actual values for $N_m$ and  $N_p$ have been fixed by some preliminary runs on the passive system
detailed in the next section. 
\section{Passive system: tuning of filament lengths}
\label{sect:Passive}
We performed some preliminary simulations to select the best suited values of the remaining 
parameters, $N_p$ and $N_m$ in the limit in which the motors are switched off (passive system). 
We observe that the total number of interaction sites, $N$, is limited by our present computer 
capabilities, mainly due to the need of vast statistics. 
We first consider a system with $N_p=250$ polymers (a safe trade between the need for 
minimizing finite size effects and available computational resources) and values for $N_m$ 
ranging from 5 to 101, to select the optimal polymer size.

The wanted number density $\hat\rho$=1 is set by appropriately scaling the simulation box
size. The target temperature, $T=0.8$, corresponds to a safe liquid state. 
This choice of density and temperature has two important consequences.
First, the passive system is in a liquid state characterized by relaxation times that are much
shorter than the typical simulation total time scale and it is comfortably in equilibrium 
with its environment. Therefore, averages of quantities that depend on only one time 
(after preparation) are time-independent and two-time correlations depend on the time-delay 
between the measurement of the two observables. More precisely, the dynamics are stationary. 
Second, the chosen thermodynamical point is well inside the stable liquid phase for the 
passive system. This  allows us, in the active case, to safely assign any non-conventional 
structural or dynamical feature to the motor activity alone and not to some ``normal'' effect 
due to the interaction with some phase boundary.
\subsection{Collective static structure}
\label{subsec:collecive}
\begin{figure}[h]
\vspace{0.15cm}
\includegraphics[width=0.49\textwidth]{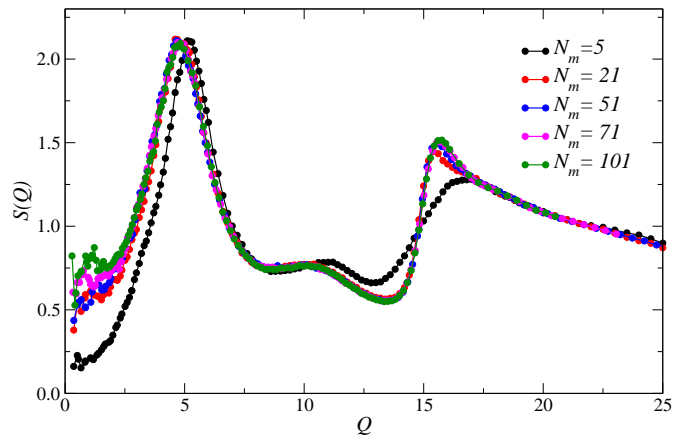}
\caption{ 
Study of the passive system at $T=0.8$ and $\hat\rho=1$.
The plot shows the collective static structure factor $S(q)$ for systems with $N_p=250$ 
polymers with length $N_m$ as indicated in the legend. For $N_m$ larger than 21 no significant 
differences in the static structure on the entirely investigated wave-vector $q$-range are observed.}
\label{fig:sq-vs-nm}
\end{figure}

We start by quantifying the polymer size dependence of the structure of the system. 
In Fig.~\ref{fig:sq-vs-nm} we show the static structure factor
\begin{equation}
S({\bf q}) = N_m^{-1} N_p^{-1} \langle \, 
\rho({\bf q},t) \rho^*({\bf q},t) \, \rangle,
\label{eq:structure}
\end{equation}
with $\rho({\bf q},t)= \sum_{i=1}^{N_m} \sum_{a=1}^{N_p} e^{i {\bf q}
\cdot  {\bf r}_{ia}(t)}$ the Fourier transform of the instantaneous density. 
${\bf r}_{ia}(t)$ is the time-dependent position of monomer $i$ in polymer $a$.  
Isotropy implies $S({\bf q})=S(q)$ and a spherical average over wave vectors of modulus $q$ is 
considered. While the curve for $N_m=5$ clearly falls away from the remaining data, the static 
structure depends only weakly on $N_m$ as shown by the collapse of all curves 
with $N_m\ge$ 21. 

A detailed description of these results goes beyond the goal of this
work, nevertheless we briefly discuss our findings in what
follows. The $S(q)$ of Fig.~\ref{fig:sq-vs-nm} show the typical
features of a polymer melt, namely the presence of two principal
peaks. The first one is usually related to inter-chains average distance
and is at $q_0\simeq 4.8$. The length associated to this value is
about 1.26 which, as we will see below, is the typical distance
between monomers pertaining to nearest neighbor polymer chains.  The
second peak, at position $q_1\simeq 15.7$ is associated to the
equilibrium bond distance $r_0$.  The feature of small intensity
located at about $q=10$ is not associated to any evident length scale,
being probably due to the convolution of density fluctuations of wave
vectors between $q_0$ and $q_1$.  Subsequent peaks, not shown in the
figure, have less intensity reflecting the loss of spatial correlation
in the liquid~\cite{Hansen}.

We note that the general shape of our curve is very similar to the one found in a united atoms model 
of 1,4-polybutadiene (see Ref.~\cite{paul04} and references therein). In contrast, and as expected, 
$S(q)$ is quite different from  the one for highly flexible polymers modeled by a 
bond-spring model with Lennard-Jones interactions complemented by the finitely extensible 
non-linear elastic potential~\cite{Chong}. 

>From this preliminary analysis $N_m=21$ appears as the 
minimal polymer length to be chosen to avoid any important dependence of the collective structure
on the size of the polymer.
\subsection{Single polymer structure}
\begin{figure}[h]
\vspace{0.2cm}
\includegraphics[width=0.49\textwidth]{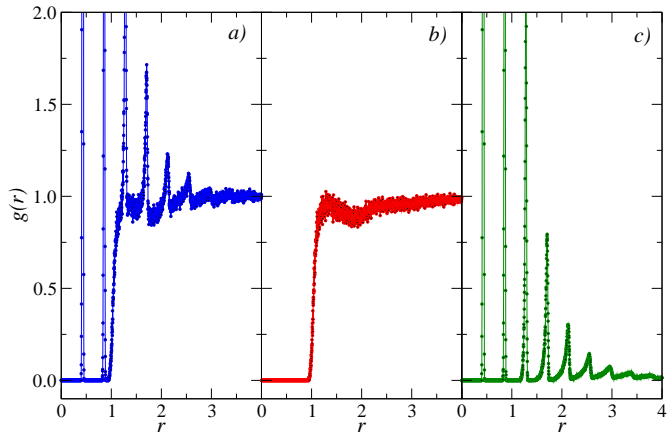}
\caption{ Study of the passive system at $T=0.8$ and $\hat\rho=1$ with
$N_p=250$ and $N_m=21$.  The pair distribution functions are computed summing over~a) 
all beads in the system;~b) beads pertaining to different polymers;~c) beads pertaining to the same polymer. }
\label{fig:sg-vs-nm} 
\end{figure}

An equivalent characterization of the static structure of the passive
system is based on the study of the pair distribution function
$g(r)=V/n^2\langle\sum_{\alpha=1}^n\sum_{\beta\ne\alpha} \delta(r-r_{\alpha\beta})\rangle$ with different
choices of the sum ranges. This allows one to disentangle the inter and intra polymer 
correlations, as shown in Fig.~\ref{fig:sg-vs-nm} for the case $N_p=250$ and $N_m=21$. 
Data in panel~a) are calculated summing over all monomers in the system; data in panels~b) and~c), instead, correspond to the sums performed over monomers pertaining to different
polymers and monomers pertaining to the same chain, respectively. Data
in~a) are the convolution of the curves shown in~b) and~c). These results give additional 
support to the description proposed in Sect.~\ref{subsec:collecive}.

The analysis of the fluctuating radius of gyration~\cite{degennes79} averaged over
all polymers
\begin{equation}
R_g^2(t)=\langle \frac{1}{N_p} \sum_{a=1}^{N_p}
\frac{1}{2N^2_m} 
\sum_{i\ne j}  |{\bf r}_{ia}(t)-{\bf r}_{ja}(t)|^2 \rangle,
\label{eq:radius-gyration}
\end{equation}
confirms that the semi-flexible chains in the melt are non-Gaussian, 
in opposition to what is predicted by the standard completely flexible polymer 
theory~\cite{doi87,degennes79}. This is shown in Fig.~\ref{fig:gyration-vs-nm}~b), 
again for the case $N_m=21$, where we plot the distribution of the radius of gyration 
for passive and active systems. The average value for the passive system, 
corresponding to $R_g\simeq 2.54$ [Fig.~\ref{fig:gyration-vs-nm}~a)], has to be 
compared with the Gaussian model value, $R_g=r_0\sqrt{N_m/6}\simeq 0.75$.
\subsection{Equilibrium dynamics}
\begin{figure}[h]
\vspace{0.25cm}
\includegraphics[width=0.49\textwidth]{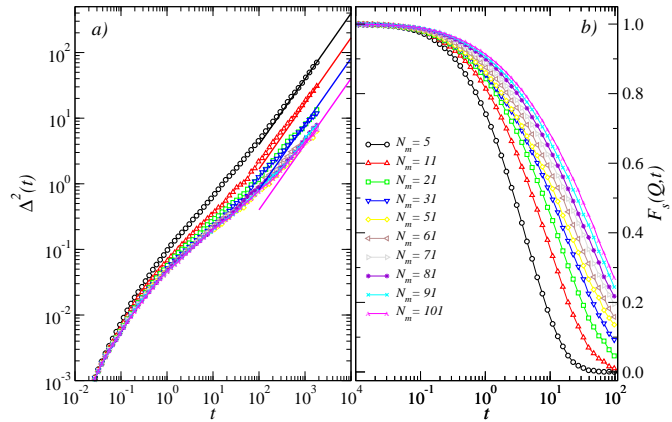}
\caption{~a) Mean-square displacement and~b) self intermediate
scattering function in the passive system with $\hat\rho=1$ at $T=0.8$ 
as a function of $t$ for different values of $N_m$ given in the legend of panel~b). 
In all cases $N_p=250$ and all monomers have been taken into account. The limited time range 
used still allows for the determination of a self diffusion coefficient from a linear
fit of the long time part of the mean squared displacement. 
Some examples are shown with solid lines. The self
intermediate correlation function is calculated at a value of $q=4.8$ corresponding to the
first diffraction peak of the static structure factor.
}
\label{fig:dyn-fqt-nm0} 
\end{figure}

\begin{figure}[h]
\vspace{0.15cm}
\includegraphics[width=0.49\textwidth]{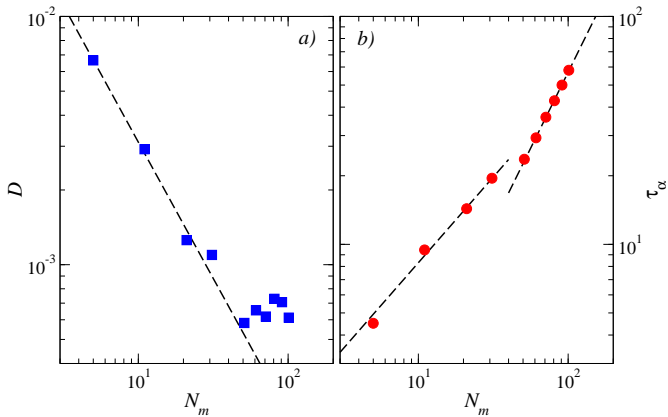}
\caption{~a) Self diffusion coefficient $D$ calculated by a linear fit
of the long time linear behavior of the mean squared displacement, as explained in 
the text. The dashed line is a guide-to-the-eye. b) The relaxation time
extracted from $F_s(q,t)$ for $q$=4.8 as described in the text. Note the apparent
dynamical crossover at $N_m^*\simeq 50$.
We remind that all quantities are calculated from all monomer coordinates, as discussed 
in the text.}
\label{fig:dyn-fqt-nm} 
\end{figure}

The question of the role played by rigidity in the dynamics of a concentrated melt of 
linear polymers is highly non-trivial (see, among others, Ref.~\cite{bulacu07}) and a 
comprehensive study of this issue goes far beyond our present interests. 
Here we are primarily interested in a qualitative characterization of the self dynamics 
of the passive polymer system. In particular, we need to check if the value $N_m=21$ which 
appears to be the best choice for a set of $N_p=250$ interacting filaments is characterized 
by a structural dynamics taking place on reasonable time scales. 
Our main results are summarized in Figs.~\ref{fig:dyn-fqt-nm0} and~\ref{fig:dyn-fqt-nm}.

In Fig.~\ref{fig:dyn-fqt-nm0}~a) we show the averaged (three-dimensional) mean-squared 
displacements 
\begin{equation} 
\Delta^2(t) =\frac{1}{N} \langle   \sum_{a=1}^{N_p}
\sum_{i=1}^{N_m} |{\bf r}_{ia}(t)-{\bf r}_{ia}(0)|^2\rangle ,
\label{eq:msd}
\end{equation}
and in panel b) we show the self-intermediate scattering functions
\begin{equation}
F_s(q,t) =\frac{1}{N} \langle \sum_{a=1}^{N_p}
\sum_{i=1}^{N_m} e^{i{\bf q} \cdot [{\bf r}_{ia}(t) - {\bf r}_{ia}(0) ]} \rangle ,
\label{eq:fqt}
\end{equation}
at $q=4.8$, the wave-vector of the first diffraction peak 
in the static structure factor, in both cases calculated considering all beads and 
as a function of $t$. 
The chain lengths  are indicated in the figure. Our data are consistent, as expected, 
with a picture where the larger the polymer size, the slower the relaxation. We also note
that these data show the typical signatures of a liquid state, with no dynamical arrest
nor crystallization.

The mean-square displacement in Fig.~\ref{fig:dyn-fqt-nm0}~a), presents the typical bending 
separating short time-scales ($t \stackrel{<}{\sim} 10^{-1}$) with ballistic motion from long 
time-scales ($t\stackrel{>}{\sim} 10^{-1}$) with, at least for short chains, diffusive 
behavior. Indeed, although the investigated time range is quite limited, for
$N_m\le N_m^*=50$ we can estimate a diffusion coefficient from the Einstein relation 
$D=\Delta^2(t)/6t$ in the limit $t\rightarrow\infty$.
These data are shown in Fig.~\ref{fig:dyn-fqt-nm}~a), and points for $N_m\le N_m^*$ are consistent
with a power law $N_m^{-1}$ (dashed line). Data for $N_m > N_m^*$ are much less convincing, saturating
at a constant value. This is possibly due to the limited time scale reached, which does not 
actually allow us to discriminate between a diffusive behavior and a Rouse-like behavior 
$\propto t^{0.61}$~\cite{doi87,degennes79,Barrat-Baschnagel-Lyulin} which should 
be a signature of limited entanglement. 

The quality of our data for the intermediate scattering functions [Fig.~\ref{fig:dyn-fqt-nm0}~b)]
allows us to extract quite comfortably an estimate for the structural relaxation time $\tau_\alpha$
by using the model independent definition $F_s(q,\tau_\alpha)=1/e$. Our results are shown in  
Fig.~\ref{fig:dyn-fqt-nm}~b). We find evidence for a dynamical crossover similar to the Rouse-like 
to reptation-like one~\cite{doi87}. More precisely, we find a relaxation time $\tau_\alpha\simeq N_m^{3/4}$ 
for $N_m<N_m^*$ and $\tau_\alpha\simeq N_m^{4/3}$ for $N_m>N_m^*$, with $N_m^*\simeq 50$. 
We tentatively associate the anomalous power-law dependencies to the fact that we are dealing with 
semi-flexible polymers. The wave-vector dependence of $\tau_\alpha$ is the usual hydrodynamic 
behavior $\tau_\alpha\simeq q^{-2}$ (not shown).

We conclude our brief study of the chain-length dependence of the structure and dynamics of a 
passive semi-flexible polymer melt by stating that, surprisingly enough, to the best of our 
knowledge, this problem has not been fully investigated in the past. The $N_m^{-1}$ dependence 
of the self-diffusion coefficient is a known scaling law of quite generic polymer dynamics 
but we know of no previous numerical studies to compare with concerning the $N_m$ 
dependence 
of $\tau_\alpha$. A complete characterization of a possible dynamical crossover corresponding to 
a particular chain size is also lacking. All these matters are of limited interest in the
context of the study of $T_{eff}$ 
but call for further work. Here we conclude by fixing the final parameter $N_m=21$
corresponding to a passive system with quite standard structural and dynamical features, 
taking place on time scales comfortably reachable in numerical experiments.
\section{Active system: the effect of motor activity}
\label{sect:Results}
We now turn to the investigation of the structure and dynamics of the active semi-flexible 
polymer melt in which motors are switched on. 
We stress here that external macroscopic conditions are fixed and the only control parameter 
is the motor activity. We recall that we quantify the motor activity via a single parameter 
$f\equiv f^M/\bar  F$, {\it i.e.}, the ratio (in the range $0.1\div 1$) between the force exerted 
by the motor and the mean mechanical force, ${\bar F}\simeq 163.5$, acting on the equivalent 
passive system. We fix the activation motor time as $\tau=500$ molecular dynamics steps.
Similar to the inactive case detailed in Section~\ref{sect:Passive}, we consider 
$N_p=250$ polymers with $N_m=21$ monomers each, for a total of $N=5250$ interaction sites. 
\begin{figure}[h]
\vspace{0.15cm}
\includegraphics[width=0.49\textwidth]{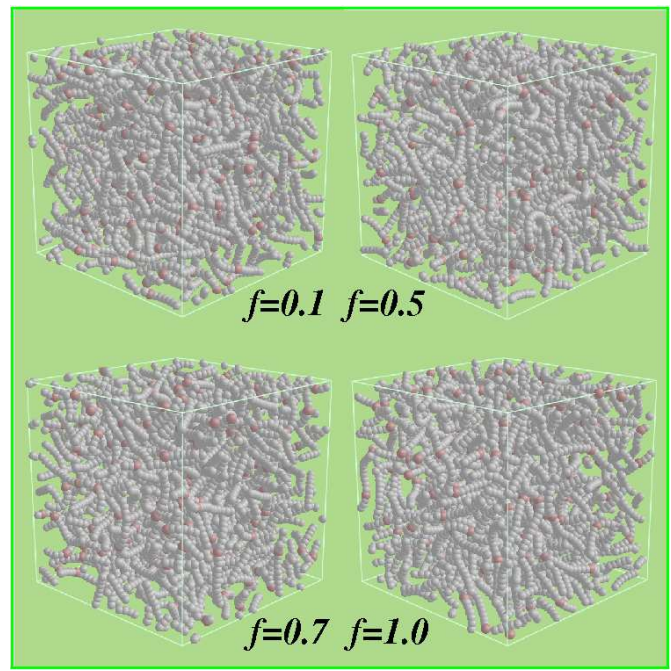}
\caption{ Typical snapshots of the system of motorized
  filaments at the indicated values of the motor activity, $f$. Brown
  beads indicate the instantaneous positions of the molecular motors.}
\label{fig:fig-confs} 
\end{figure}

In Fig.~\ref{fig:fig-confs} we show typical snapshots of the system produced following
the above prescriptions, at the indicated values of motor activity. Differences are not 
immediately evident and we need to characterize quantitatively the effect of 
motors on intramolecular and intermolecular structure.
\subsection{Thermodynamics}
\label{subsec:thermo}
\begin{figure}[h]
\vspace{0.15cm}
\includegraphics[width=0.49\textwidth]{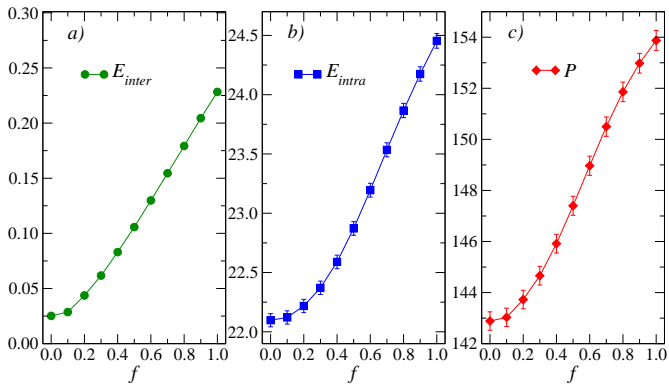}
\caption{  Thermodynamical observables for the active
  system, as a function of the motor activity $f$ defined in the
  text. From left to right we show the inter-molecular repulsive
  Lennard-Jones contribution~a), the (total) intra-molecular energy~b)
  and the total pressure~c) (defined in Eqs.~(\ref{eq:pot-inter}) and
  (\ref{eq:pot-intra}) and calculated from the usual virial relation,
  respectively).}
\label{fig:thermo-vs-f} 
\end{figure}

We start by analyzing the $f$-dependence of a few thermodynamic observables. 
Figure~\ref{fig:thermo-vs-f} presents the $f$ dependence of the inter-molecular and 
intra-molecular energy, see Eqs.~(\ref{eq:pot-inter}) and (\ref{eq:pot-intra}) respectively,  
as well as the pressure defined through the usual virial relation
$PV=N T + 1/3 \langle \sum_{ia}{\bf r}_{ia}\cdot {\bf f}_{ia} \rangle$~\cite{allen89}.
The three quantities increase smoothly for moderate forces, $f\stackrel{<}{\sim} 0.2$, next 
they cross over to an approximately linear, in the case of $U_{inter}$, and sigmoidal, 
in the cases of $U_{intra}$ and $P$, shape. 
We note that the overall variation of the intermolecular potential energy is of order 8,
while it is of order 0.1 for intramolecular (sum of the repulsive Lennard Jones and 
bonding) energy and total pressure. We also note that for the latter only the configurational
term plays a role, being $T$ constant. The results of panel~a) seem to suggest a picture
where distances between beads pertaining to different filaments decrease with increasing $f$, 
therefore indicating crowding. 

The pressure results are also very interesting. Indeed, pressure is related 
to the (negative) diagonal components of the stress tensor. Thus we can anticipate that motor 
activity has a non-trivial effect on the mechanical properties of the filaments. This is actually 
what is observed in real systems, see {\it e.g.}~Ref.~\cite{mizuno07}.
Already at the level of simple thermodynamic quantities related to configurational properties 
motor activity is non negligible and we expect to find sensible implications on the static structure
as well.
\subsection{Collective structure}
\label{subsect:structure}
In this section we consider the static structure factor $S(q)$, see Eq.~(\ref{eq:structure}), 
and the gyration ratio defined in Eq.~(\ref{eq:radius-gyration}) and we try to give a consistent picture
of the motor activity effect on the static structure of the system.

\begin{figure}[h]
\vspace{0.25cm}
\includegraphics[width=0.49\textwidth]{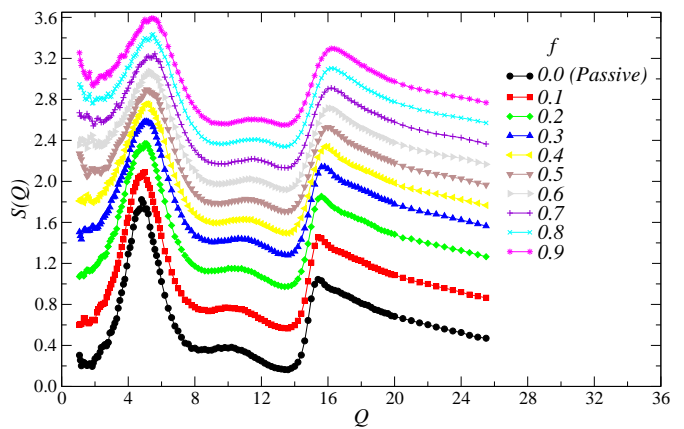}
\caption{ Static structure factor in the active sample 
at the investigated motor activities parametrized by $f$
(highest $f$ on top). Data have been shifted vertically for clarity. The first diffraction
peak increases with $f$ as described in the text.}
\label{fig:sk-vs-nm} 
\end{figure}

In Fig.~\ref{fig:sk-vs-nm} we show the static structure factor as a
function of motor activity. All curves have been shifted by a fixed
amount for clarity, $f$ increases from bottom to top.  The general
features of the structure factor are not altered by the action of the
motors but, upon careful inspection, one observes some subtle changes.

Upon increasing the strength of the motor activity, the main
peak position shifts to higher values of $q$, meaning that the average
nearest neighbor distance between beads pertaining to different
polymers decreases (recall that the first and highest peak is mainly 
determined by interchain correlations). This is also consistent with the crowding
picture suggested from the study of the thermodynamic quantities 
in Sect.~\ref{subsec:thermo}. The width of the first peak increases while 
its height decreases with increasing $f$. 
This trend is similar to the one that one finds
when increasing the bath temperature in a passive polymer liquid, see
for instance the simulations in~\cite{Chong}. The motors have then the mixed
effect of making the melt more compact and more disordered simultaneously.

\begin{figure}[t]
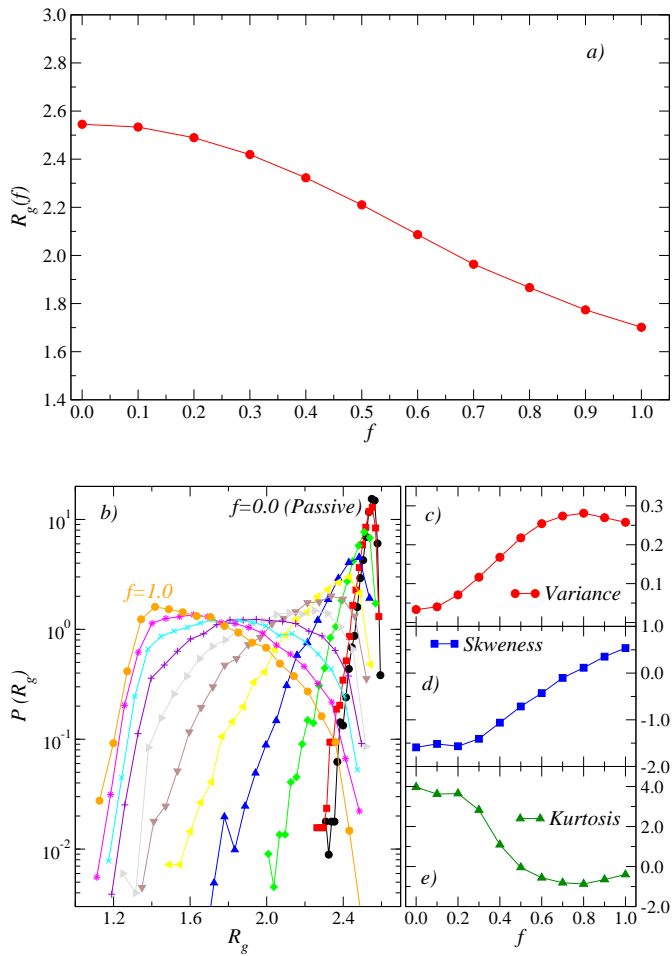

\vspace{0.15cm}
\includegraphics[width=0.49\textwidth]{./average-rg.eps}
\vspace{0.15cm}

\includegraphics[width=0.49\textwidth]{./gyration.eps}
\caption{~a)~Averaged gyration ratio.~b) Probability
  distributions of the gyration radius at all investigated motor
  activities. The structure is highly non-trivial with an average
  value shifting towards lower values for increasing $f$. A more
  precise characterization of the distribution is given by its
  variance~c), skewness~d) and kurtosis~e).}
\label{fig:gyration-vs-nm} 
\end{figure}

The second and third peaks in $S(q)$ follow the same trend, see
Fig.~\ref{fig:sk-vs-nm}, suggesting that the effect of motor activity
is to fold the chains. This is consistent with the fact that the
radius of gyration decreases with increasing $f$, see
Fig.~\ref{fig:gyration-vs-nm}~a). 

Working in real space one corroborates this picture. Not much information can be 
extracted by simply looking at the system snapshots of Fig.~\ref{fig:fig-confs}. 
In particular, the fact that  increasing $f$ has a sensible 
influence on the gyration ratio of the filaments is not evident from the snapshots, as
it is when we quantify the analysis in
 Fig.~\ref{fig:gyration-vs-nm}. We show in Fig.~\ref{fig:gyration-vs-nm}~a) 
the average value of the gyration ratio as a function of motor activity. 
It decreases from $2.54$ for the passive system to $1.7$ for $f=1$. 

The probability distribution
of $R_g$ [Fig.~\ref{fig:gyration-vs-nm}~b)] suggests 
an even more complex scenario. In the passive system it is non-Gaussian and 
very asymmetric. We can characterize these curves very
precisely with the variance, skewness and kurtosis~\cite{momenta-note} 
dependence on $f$, as shown in Fig.~\ref{fig:gyration-vs-nm}~c)-e). 
A cross-over between regions of different curvature is evident, but we are not 
in the position to justify this complex behavior. It would be very interesting 
to compare these data to analogous experimental measurements using
scattering techniques.

In conclusion, motor activity apparently has the effect of pushing
closer the filaments, which at the same time fold substantially. There
is a competition between two effects: crowding, which is related to
excluded volume effects and tends to slow down the dynamics; and
folding of the chains, which dynamically lifts the topological
constraints, therefore decreasing entanglement. This must have
important consequences on the long-time evolution of our system, as we
will see in the next section.
\subsection{Dynamics}
\label{subsect:dynamics}
In this section we quantify the effect of motor activity on the
dynamics and we compare it to the relaxation in the passive limit. We
stress again that we work at constant volume and temperature, which
means that all observed differences in the dynamical behavior are only
due to the motor effects which are in any case chosen to be mild.

\begin{figure}[h]
\vspace{0.25cm}
\includegraphics[width=0.48\textwidth]{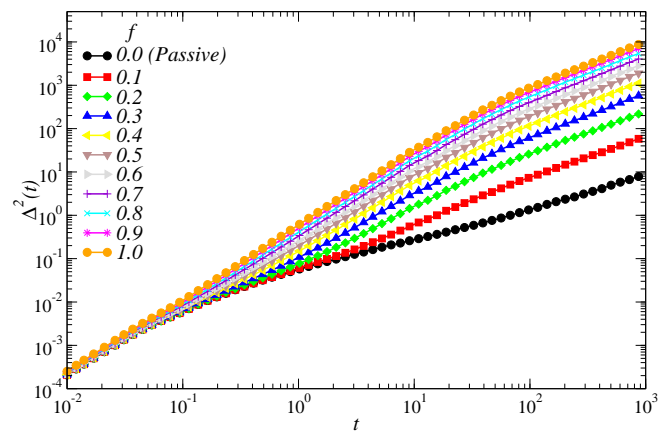}
\caption{ Mean-squared displacement in the active
system with the motor activities given in the key (higher activities
on top).  }
\label{fig:msd-vs-f} 
\end{figure}

\begin{figure}[t]
\vspace{0.15cm}
\includegraphics[width=0.48\textwidth]{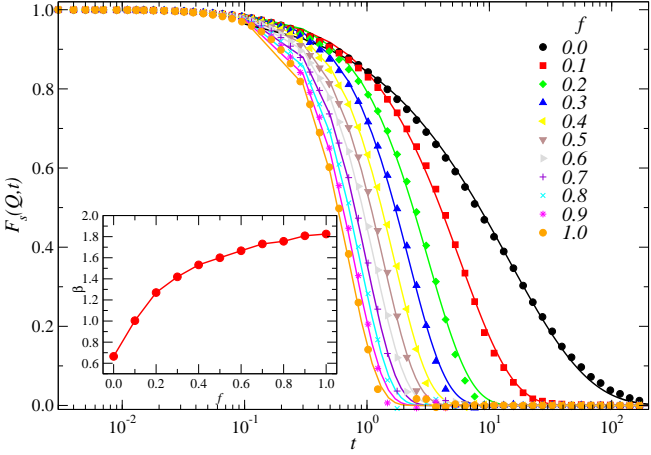}
\caption{ Main panel: Time-dependence of the intermediate 
scattering functions at the wave-vector corresponding to the first maximum 
of the static structure factor at the investigated motor activities 
(activities increase to the left). Inset: values of the exponential parameter 
$beta$ estimated from a fit of the data to the form $\exp-(t/\tau)^\beta$
(see text for details).}
\label{fig:msd-fqt-vs-f} 
\end{figure}

Figures~\ref{fig:msd-vs-f} and \ref{fig:msd-fqt-vs-f} (main panel)
display the mean-square displacement, Eq.~(\ref{eq:msd}), and the
self-intermediate scattering function, Eq.~(\ref{eq:fqt}) -- for the
wave-vector $q_0\simeq 4.8$ of the main peak in $S(q)$ in the passive
case -- as a function of time-delay for all investigated motor forces
given in the key. The time range considered is quite extended and
allows us to comfortably reach the diffusive regime of the global
displacement in which the self-intermediate scattering functions
safely decay to zero.

Qualitatively, the data resemble the ones found in the passive system, 
see Fig.~\ref{fig:dyn-fqt-nm0}, but depend now on the motor strength. 
All the data are compatible with a picture where the collective dynamics of filaments 
gets faster under stronger motors. This is also consistent with our results on the
gyration ratio and supports our interpretation: the augmented folding of the filaments 
seems to solve local topological constraints and diminishes the effect of entanglement. 

Additional interesting information is included in the data of
Fig.~\ref{fig:msd-fqt-vs-f}.  Solid lines are best fit of our data to
a curve of the form $\exp-(t/\tau)^\beta$.  Values for the $\beta$
coefficient as a function of $f$ are included in the inset.  Data
increases with $f$, implying that the relaxation crosses over from
stretched exponential ($\beta\le 1$) to compressed exponential
($\beta>1$). This unusual shape for the intermediate scattering
function has been already observed in soft matter systems, like
colloidal gels, emulsions or Laponite suspensions, as surveyed in a
number of reviews, see, {\it e.g.}, Ref.~\cite{cipelletti05}.  In these
references, the cross over is related to randomly distributed
internal stress sources acting on the sample. It is tempting to
associate this sources to motor activity but, again, it is difficult
to address this subject here.

\begin{figure}[t]
\vspace{0.15cm}
\includegraphics[width=0.49\textwidth]{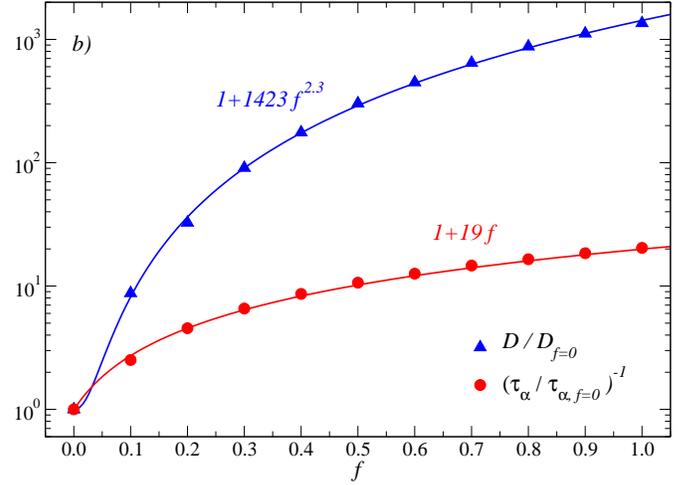}
\caption{ Diffusion coefficient $D$ and structural
(self-)relaxation time $\tau$ as a function of $f$. 
The relaxation time values have been estimated from a fit of the data to the 
form $\exp-(t/\tau)^\beta$ (see text for details).
The data follow power laws with exponents indicated in the figure.}
\label{fig:diff-tau-vs-f}
\end{figure}

An even more quantitative analysis is presented in Fig.~\ref{fig:diff-tau-vs-f}, in which 
the dependence of the diffusion coefficient and the (inverse) relaxation time (estimated by the
fit detailed above) on the 
strength of the motors is analyzed. Data are shown renormalized to the value of the 
passive system. It is very interesting to note that both sets of data are consistent with 
two very simple functional forms (solid lines in the figure), 
$D/D_{f=0}\simeq 1+1423\times f^{2.29}$ and $(\tau/\tau_{f=0})^{-1}\simeq 1+19\times f$, 
respectively. Note the extremely large variation of the diffusion coefficient that extends
over three decades, and the less pronounced variation of the inverse relaxation time.

An observation is in order at this point. The $f$-dependence of the diffusion coefficient
is reminiscent of the findings discussed in Refs.~\cite{palacci10} and~\cite{howse07}.
For the active colloidal particles used in these studies the effective diffusion 
coefficient can be written as $D=D_o+1/3 \  v^2 \tau_R$, where
$ D_o\equiv D_{f=0}=k_B T/(6\pi\eta R)$ is the diffusion coefficient of the passive system,
$v$ is the velocity of the particle and $ \tau_R=8\pi\eta R^3/(k_BT)$ 
is the rotational relaxation time. $R$ and $\eta$ are the typical radius of the colloidal 
particle and the viscosity of the solvent, respectively. 
This relation can also be written as $D/D_o=(1+1/9 P_e^2)$,
where $P_e$ is the Peclet number, defined as $P_e=vR/D_o$
and characterizes the particle activity~\cite{palacci10,baskaran09}.

In our case the exponent ($2.3\pm 0.1$) is close to $2$. The
difference could be ascribed to the uncertainties in the determination
of $D$ or to the presence of corrections to the correct scaling.  It
is tempting to imagine that $f$ is in this case playing the role of
the Peclet number.  We will see in our final discussion that this idea
is indeed reasonable and our results on the effective temperature and
analogous experimental data~\cite{palacci10} seem to be compatible
with the conjecture.

We have not been able to find a rationale for the linear dependence of
the inverse of the relaxation time, which is also a quite striking
result.
\section{Effective temperature(s) in complex active matter}
\label{sect:temperatures}
The analysis above proved that both structure and dynamics of the
active system are influenced by motor activity in a very complex
fashion. We ask now whether it is possible to embed all this
complexity in a single parameter, also prone to be directly determined
in experiments. The chosen parameter is the effective temperature
$T_{eff}$ a quantity already well characterized in different contexts,
as we have discussed in the introduction~\cite{cugliandolo-review}.

In equilibrium, a model independent relation exists between the linear
response of an observable, say $A$, to a perturbation applied to the
system and the correlation between the same observable and the one
that ``receives" the perturbation, say $B$.  This is the
fluctuation-dissipation theorem (FDT) that reads
\begin{equation}
\chi_{AB}(t_1,t_2) = \frac{1}{T} [C_{AB}(t_1,t_1)-C_{AB}(t_1,t_2)],
\label{eq:FDT1}
\end{equation}
where 
\begin{eqnarray}
\chi_{AB}(t_1,t_2) &=& 
\left. \int_{t_2}^{t_1} dt' \ \frac{\delta \langle A(t_1)\rangle_h}{\delta h_B(t')} \right|_{h_B=0},
\nonumber\\
C_{AB}(t_1,t_2) &=& \langle A(t_1) B(t_2) \rangle,
\end{eqnarray}
and the perturbation is such that $H\to H-h_B B$. The prefactor $T^{-1}$ is 
the inverse of the temperature of the bath with which the system is in contact. 
In equilibrium $C_{AB}(t_1,t_1)$ is a constant, both $\chi_{AB}(t_1,t_2)$ and 
$C_{AB}(t_1,t_2)$ are just functions of the time difference $t=t_1-t_2$, 
and Eq.~(\ref{eq:FDT1}) becomes 
\begin{equation}
\chi_{AB}(t) = \frac{1}{T} [C_{AB}(0)-C_{AB}(t)] \; , 
\label{eq:FDT2}
\end{equation}
for $t>0$.
In the frequency domain the FDT reads $2T \Im R_{AB}(\nu)=\nu C_{AB}(\nu)$ 
with the sign convention for the Fourier transform such that
$R_{AB}(\nu)=\int_{\-\infty}^\infty dt \ e^{i\nu t} R_{AB}(t)$. 
\begin{figure}[t]
\vspace{0.25cm}
\includegraphics[width=0.49\textwidth]{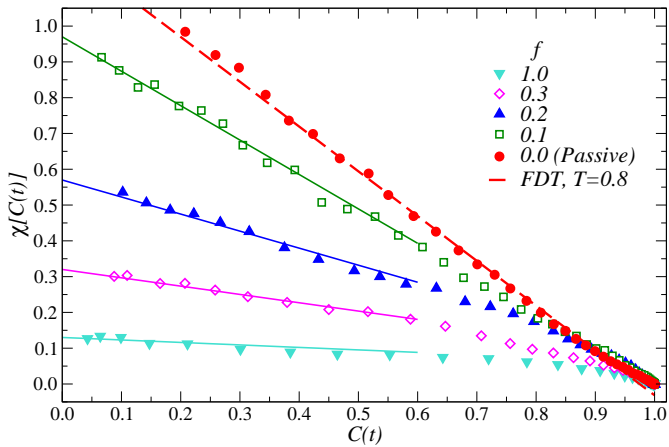}
\caption{ Fluctuation-Dissipation relations in the
motorized polymer melt.  The red data on top correspond to the
passive system. Solid lines are linear fits used to estimate the
value of the effective temperatures. The dashed (red) line has slope $-1/T$,
thus confirming the validity of the FDT at equilibrium for the passive system. }
\label{fig:R-C-vs-f} 
\end{figure}

\subsection{Correlation-response parametric plot }
One way of measuring the effective temperature in a non-equilibrium
steady state consists in using the parametric relation between the
out-of-equilibrium integrated linear response, $\chi_{AB}(t)$, and
the correlation of a chosen pair of observables, $C_{AB}(t)$, and
associating minus its inverse slope with a possibly time-dependent 
parameter
$T_{\sc eff}(t)$~\cite{cugliandolo97,cugliandolo-review}:
\begin{equation}
\chi_{AB}(t) = \frac{1}{T_{\sc eff}(t)} [C_{AB}(0)-C_{AB}(t)]
\; . 
\end{equation}
Such measurements were performed in different {\it aging}, that is to say non-stationary, 
glassy systems (for details see~\cite{glassy-Teff}) and in out of equilibrium steady states 
(see, among others, Refs.~\cite{Berthier-Barrat,Kolton,Makse}) demonstrating that
$T_{eff}(t)$ is piecewise constant taking, typically, two values: the one of the 
environmental bath at short time-delays and another one characterizing the structural 
relaxation at long time-delays.

In interacting particle problems~\cite{Berthier-Barrat} it is
customary to use $A(t) =1/N \sum_{i=1}^N\epsilon_i e^{i\vec q\cdot
  \vec r_i(t)}$, and $B(t) = 2 \sum_{i=1}^N \epsilon_i \cos[\vec
  q\cdot \vec r_i(t)] $, where the field $\epsilon_i=\pm 1$ with
probability a half. With this choice the relevant equilibrium
correlation function is the intermediate scattering function of
Eq.~(\ref{eq:fqt}) once the results are averaged over many instances
of the $\epsilon_i$. We have also used these observables with $q=q_0$
and we averaged over $10^2$ independent field realizations.

We performed these measurements for different values of the forcing $f$ and the results are shown 
in Fig.~\ref{fig:R-C-vs-f} in the form of the parametric plot $\chi(t)$ against $C(t)$. The
curves clearly show two regimes with a relatively smooth crossover between them. We do not attribute 
particular importance to the first regime, corresponding to small $t$ or high frequencies,
which represents the fast vibration of monomers, surrounded by their environment, before any structural 
relation takes place.  The second regime, being the one of long $t$ and small frequencies, is 
more interesting since it describes the actual structural relaxation in the sample.  

In all cases, the parametric plot in this regime is rather well described by a straight line~\cite{note}, 
implying that $T_{\sc eff}$ is a constant over this time-scale. In the figure the numerical data are 
represented with points and the lines are the result of fits to linear functions performed on the
long-times part of the curves. For the passive system the construction yields a straight line with 
slope equal to $-1/T$, confirming that the FDT holds in this case. The active systems have
decreasing absolute value of the slope for increasing forcing and will be compared to other measurements 
of $T_{\sc eff}$ below. Data from the $\chi(C)$ determination of $T_{eff}$
are reported in Fig.~\ref{fig:Teff-final} with open (red) squares
where we also confront to other measurements  as we will discuss in depth later.
\subsection{Tracer particles}
\begin{figure}[h]
\vspace{0.15cm}
\includegraphics[width=0.45\textwidth]{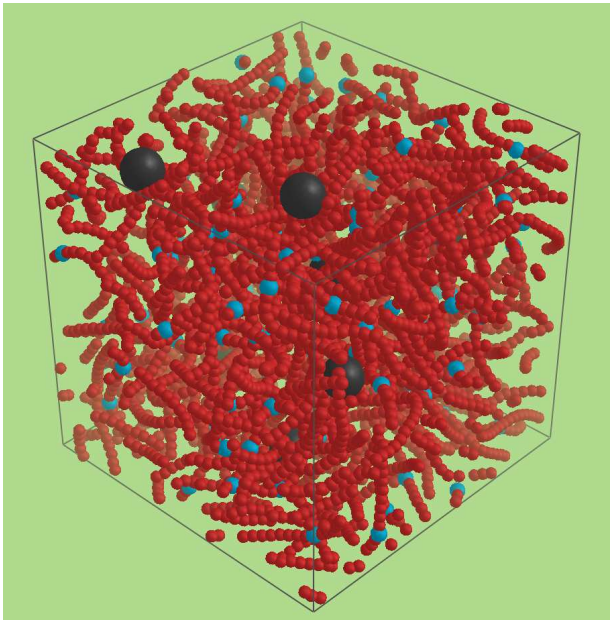}
\caption{ Molecular dynamics computer simulation
  implementation of the experiment with tracer particles. We show a
  typical polymer configuration with $10$ tracer particles; $50$ per
  cent of polymers are motorized. Tracers have been drawn with an
  augmented radius for clarity.}
\label{fig:conf-poly-trac} 
\end{figure}

The construction detailed in the previous section is very powerful and provides
us with direct unambiguous results. Indeed, it is at the same time a tool to 
{\em demonstrate} the existence of an effective temperature and an operative 
method to {\em calculate} it~\cite{Berthier-Barrat}. The method is easily 
exploitable in computer simulations, 
where the microscopic implementation of the construction
does not pose particular problems. The situation is completely different in experiments,
where the coupling of each (complex) unit forming the system to the perturbation 
field is far from obvious. This difficulty has been circumvented by considering the dynamics 
of {\em  tracers}  (or {\em intruders}) of simple geometry, {\it e.g.}, spheres. 
These objects of micrometric size are nowadays quite simple to control and manipulate by 
active microrheology techniques. When immersed in extended complex environments they have 
been demonstrated to couple to them as to an external reservoir. Their dynamics 
provide a wealth of information related to the underlying matrix state.
Below we discuss a few methods to implement these ideas in our simulations, 
based on the use of {\em free} and {\em driven} tracer particles, respectively. 
A typical snapshot of our filament system with embedded tracer particles 
is shown in Fig.~\ref{fig:conf-poly-trac}. We have used spherical particles that, 
for visualization convenience, we draw with an augmented radius in the figure.

\subsubsection{Free tracers}
The effective temperature should be measurable with a fine tuned
thermometer that couples to the structural relaxation of the sample.
A possible realization of an adapted thermometer is constituted by a
{\em massive} -- {\it i.e.}, of mass much larger than the mass of the
polymer beads -- free tracer particle immersed in the system. Indeed,
it is possible to show that the kinetic energy of the tracer can be
related to the system's effective temperature via an equipartition
theorem for slow modes.  This kind of measurement were suggested
in~\cite{cugliandolo97} and they were performed experimentally in
granular matter~\cite{Makse-granular}, colloidal
suspensions~\cite{Teff-exp-coll} and numerically in models of granular
matter~\cite{Makse} and sheared atomic Lennard-Jones
mixtures~\cite{Berthier-Barrat}.

We consider $N_{tr}=10$ tracer spherical particles of variable mass $m_{tr}$ in the interval 
$[10-10^5]$, which interact with all monomers in the sample via the repulsive Lennard-Jones 
potential of Eq.~(\ref{eq:pot-inter}). Tracers are not coupled to any external thermal bath 
and their dynamical evolution is controlled by the usual Newton equations of motion that we 
have integrated with the velocity-Verlet method with $\delta t=0.001$ in the NVE ensemble. 
The dynamics of polymer beads is still treated by a Langevin approach. 
Note that the tracers do not interact among themselves and the use of $N_{tr}$ tracers 
allows for better statistics. 

\begin{figure}[h]
\vspace{0.15cm}
\includegraphics[width=0.49\textwidth]{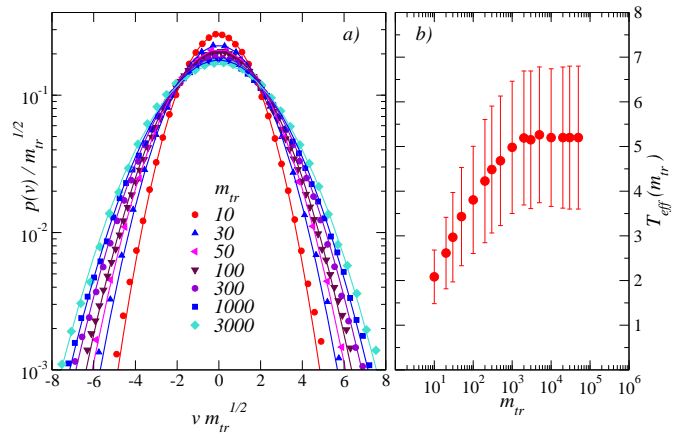}
\caption{
~a) Rescaled probability distribution of the
velocities of the massive tracer particles immersed in an active 
sample with $f=0.5$. The distributions are remarkably Gaussian for all tracer masses. 
Their variance increases with increasing mass.~b) The effective temperature, extracted from
a Maxwell-like fit to the distributions shown in panel~a),
as a function of the tracers' mass $m_{tr}$.}
\label{fig:results-poly-massive}
\end{figure}

The results of our measurements for $f=0.5 $ are shown in Fig.~\ref{fig:results-poly-massive}. 
Panel~a) displays the probability distributions of the
tracer velocity, $p(v)$, where $v=|\mathbf{v}|$ and its dependence on the tracer's mass. 
The data are shown using the scaled variables $p(v)/m^{1/2}_{\sc tr}$ against $v m^{1/2}_{tr}$. 
The distributions are very close to Gaussian for all masses. The data are shown with points and 
the lines are fits to this functional form from which we extract an estimate for the effective 
temperature as a function of the tracer's mass that we show in panel~b). 

The interpretation of these results is quite clear. Light tracers quickly react to the rapid 
motion of the polymers and test the short time-delay dynamics of the sample. In contrast, 
heavy tracers are insensitive to such rapid motion and displace themselves only due to the 
large-scale structural reorganization of the sample, that is to say, the long time-delay 
dynamics of the embedding matrix. The interesting long times -- or low frequency -- value  
of the effective temperature is then accessed in the large mass limit of this measurement. 

>From the figure we see that at $m_{tr}\simeq 10^3$ the variation of $T_{\sc eff}$ 
stops and the curve saturates to a constant level from which we read 
$T_{\sc eff}(f=0.5)\simeq 5.1$. Similar measurements for other values of the motors strength 
yield the data collected in Fig.~\ref{fig:Teff-final} with open (blue) circles.
In what follows we will consider $m_{tr}=10^3$ which corresponds to the smallest tracers
mass coupling to the relevant slow modes of the filaments.
\begin{figure}[t]
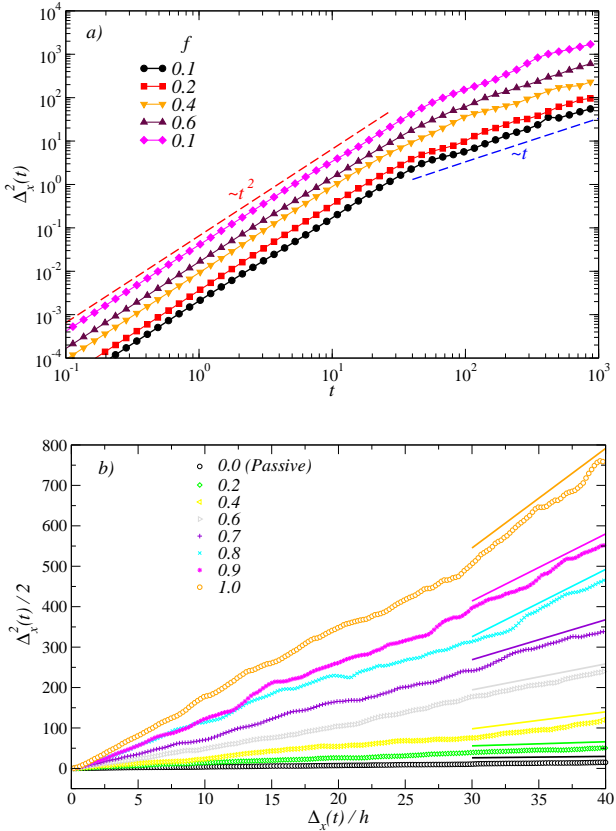

\vspace{0.15cm}
\includegraphics[width=0.45\textwidth]{./msd-tracers.eps}
\vspace{0.5cm}

\includegraphics[width=0.45\textwidth]{./fig-5.21.eps}
\caption{~a) Uni-dimensional mean-squared displacement of the free tracer particles 
at different values of the motor activity.~b) Parametric plot of the tracer particles 
mean squared displacement as a function of the driven displacement, 
at the indicated values of motor activity. The effective temperatures have been 
estimated with a linear fit (solid lines) to the large mobility part of the curves. 
}

\label{fig:diff-vs-mobility-vs-f-trac} 
\end{figure}
\subsubsection{Driven tracers}
Yet another independent measurement of $T_{\sc eff}$ is based on the use of driven 
tracer particles. More precisely, it is given by the study of diffusion and mobility
of an ensemble of non-interacting tracer particles immersed in the active sample.
These two quantities are related by the Einstein relation, one particular form of the
fluctuation dissipation theorem. The (free) mean squared displacement, 
is the relevant equilibrium correlation function; the associated response function is 
the displacement induced by applying a small constant external force of intensity $h$ 
to the tracer. Due to isotropy, we focus on the one dimensional version of these
observables and consider the external force oriented in the $\hat{x}$ direction.

In Fig.~\ref{fig:diff-vs-mobility-vs-f-trac}~a) we show the one dimensional free 
mean-square displacement for tracer particles, 
$\Delta^2_x(t)=1/N_{tr}\langle\sum_{i=1}^{N_{tr}}(x_i^{tr}(t)-x_i^{tr}(0))^2\rangle$, 
in the $\hat{x}$ direction as a function of time-lag, for different forcing strengths 
given in the key. Results in the other two directions are equivalent. In all cases the 
tracer particles mean-square displacements cross over from ballistic motion at short 
time-delays to diffusive motion at longer time-delays. These two limits are shown with 
straight lines as guides-to-the-eye in the figure. 

The one-dimensional mobility in the $\hat{x}$ direction is calculated as 
$\chi_x(t)=\Delta_x(t)/(ht)$, where 
$\Delta_x(t)=1/N_{tr}\langle\sum_{i=1}^{N_{tr}}(x_i^{tr}(t)-x_i^{tr}(0))\rangle$
is the displacement calculated in independent simulations by applying to the tracers 
a small force $h=0.1 \bar{F}\hat{x}\simeq 16.5\hat{x}$. 

Conforming to the Einstein relation $D_x/\chi_x=T$, with $D_c$ the diffusion coefficient in the 
$x$ direction and $\chi_x$ the saturated mobility, the effective temperature should 
also be the parameter linking the tracers' mean-square displacement and their driven
displacement under a weak external applied force:
\begin{equation}
\frac{\Delta^2_x(t)}{2} = T_{eff}\frac{\Delta_x(t)}{h}
\label{eq:einstein-relation}
\end{equation}
in the long time-lag limit in which the dynamics is diffusive.
This equation is plotted in parametric form in Fig.~\ref{fig:diff-vs-mobility-vs-f-trac}~b) 
for the passive system and seven motor strengths given in the key. 
The effective temperature is given by the slope of these curves, 
as shown in the figure. These data are included in Fig.~\ref{fig:Teff-final} 
as open (green) triangles.
\begin{figure}[h]
\vspace{0.25cm}
\includegraphics[width=0.49\textwidth]{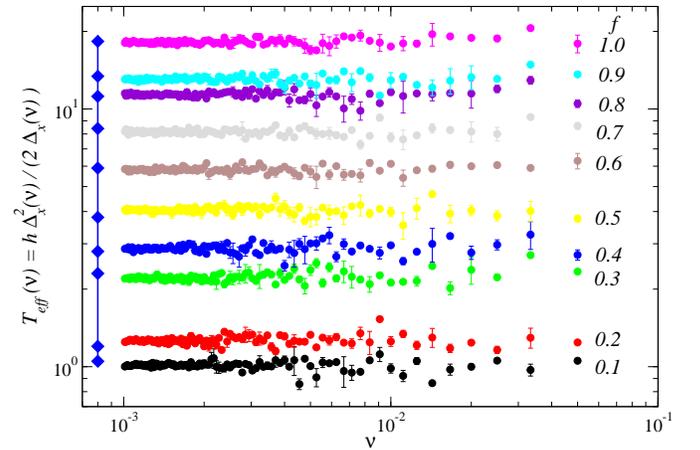}
\caption{ 
The effective temperature, defined as the ratio $h\Delta^2_x(\nu)/(2\Delta_x(\nu))$, 
as a function of frequency. We represent as error bars the imaginary part of the ratio,
to quantify the precision of the numerical Fourier transformation.
Diamonds on the left are the data extracted from the linear fit shown in 
Fig.~\ref{fig:diff-vs-mobility-vs-f-trac} b).
}
\label{fig:teff-vs-freq} 
\end{figure}
\subsubsection{Frequency-dependent $T_{eff}$.} 
Parametric plots in the time-domain 
(Figs.~\ref{fig:R-C-vs-f} and \ref{fig:diff-vs-mobility-vs-f-trac} b)) are the standard way 
of identifying the values taken by the effective temperature in well-separated dynamic 
regimes each with a constant value of $T_{eff}$. They may hide, though, long crossovers 
in which $T_{eff}$ does indeed depend on time. Besides, the determination of an effective 
temperature in experiments very often relies on measurements of spontaneous and induced 
fluctuations in the frequency domain.

In Fig.~\ref{fig:teff-vs-freq} we represent the same data shown in 
Fig.~\ref{fig:diff-vs-mobility-vs-f-trac}~b) in a closer way to what is implemented 
in experiments. We have Fourier transformed both the tracers' mean-square displacement and 
displacement under an external field and we present the ratio,
\begin{equation}
T_{eff}(\nu) = \frac{h}{2}\frac{\Delta^2_x(\nu)}{\Delta_x(\nu)}, 
\end{equation}
as a function of frequency. This determination of $T_{eff}$ should 
coincide with the one of the time-domain in the regimes in which $T_{eff}$ is constant. 
And this is indeed what we have found: for all the activation strengths and in the accessible 
frequency range, the curves show a relatively constant force-dependent plateau. 
In the limit of very low frequency we recover the frequency-independent values 
(diamonds) obtained by the linear fit of the parametric plot 
in the time-domain (Fig.~\ref{fig:diff-vs-mobility-vs-f-trac}).
Although the accessible frequency range is limited by the sampling rate (upper bound)
and the total time length (lower bound) of the trajectory and in principle one 
cannot exclude a different behavior outside this region,
we conclude that no evidence of frequency dependent effective temperature
has been found in the present system. 
\subsubsection{Free tracer dynamics.} 
\begin{figure}[t]
\vspace{0.4cm}

\includegraphics[width=0.49\textwidth]{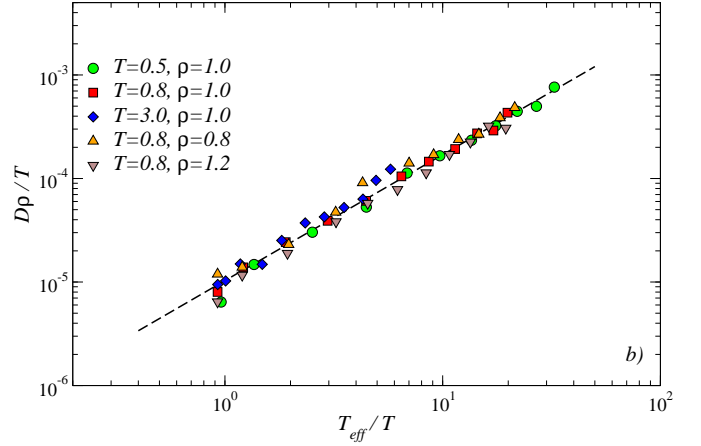}
\caption{ Parametric plot of the rescaled diffusion
  coefficients for a free tracer at different values of the number
  density $\rho$ and bath temperature $T$ as a function of the
  rescaled effective temperatures.  The dashed line is a linear fit to
  the data. More details are given in the text.}
\label{fig:dyn-vs-nm} 
\end{figure}

The power of methods based on intruders has been largely demonstrated
in the literature~\cite{gradenigo10} and in the present work. The
results contained in Fig.~\ref{fig:dyn-vs-nm} confirm this fact and
also allow one to add an interesting information: knowledge of the
effective temperature dependence on $f$ in {\em particular}
macroscopic external conditions, {\it i.e.}, temperature or pressure
(density), allows one to predict $T_{eff}$ at any different
macroscopic external conditions with one free tracer measure only.
Indeed, we have found that a simple power law relates the diffusion
coefficient of a free tracer, $D$, to the effective temperature
characterizing the polymer matrix. All data corresponding to different
temperatures or densities can be collapsed on a master curve by a
simple rescaling:
\begin{equation}
\frac{D\rho}{T} \propto \left( \frac{T_{\sc eff}}{T}\right)^\alpha,
\label{eq:rescaling}
\end{equation}
with $\alpha\simeq 1$.
\begin{figure}[b]
\vspace{0.4cm}

\includegraphics[width=0.49\textwidth]{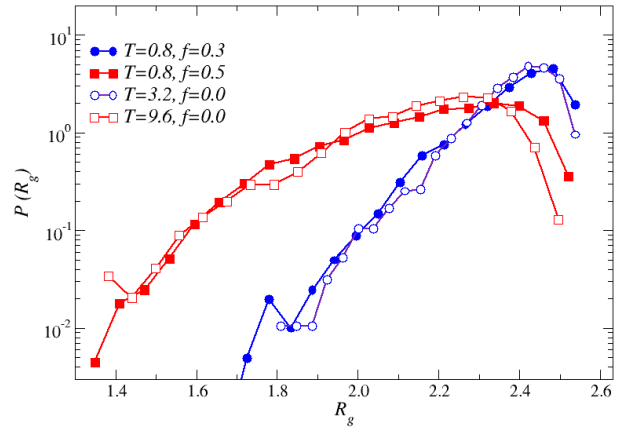}
\caption{ Comparison between gyration radius of passive
  and active systems at the indicated temperatures and motor
  activities. See discussion in the text for details.  }
\label{fig:gyration-teff} 
\end{figure}
This is exhibited in Fig.~\ref{fig:dyn-vs-nm} where we plot the diffusion coefficients
of tracer particles against the effective temperature calculated by the massive 
particle method. This means that once the master curve is known, we can predict the 
value of $T_{eff}$ by just following the free dynamics of the tracers ({\it i.e.}, by
computing $D$). 

We conclude this section with the following
observation~\cite{note-referee}.  In the case of out-of-equilibrium
glassy systems, one-time quantities converge to constants and, in
particular, thermodynamic observables receive contributions from the
fast regime at the ambient temperature and the slow regime at the
(higher) value taken by the effective temperature, both as if they
were equilibrium ones. In a sense, for these observables, the
out-of-equilibrium system behaves as if it were equilibrated at
$T_{eff}$. We could therefore ask, what can we say about the structure
of the passive system when equilibrated at $T_{eff}$?  In
Fig.~\ref{fig:gyration-teff} we plot the probability distributions of
the radius of gyration (a one-time quantity), at the indicated values
of bath temperature and motor activity. It is evident that the
distributions at the higher bath temperatures for the passive case
(open symbols) nicely reproduce those at $T=0.8$ and the indicated
values of activity (closed symbols).  The problem is that these bath
temperatures are not those we could predict from the data in
Fig.~\ref{fig:Teff-final}. We believe that the connection between effective
temperature and static structure is even more problematic that one could
expect.

\subsection{Survey on $T_{eff}$ vs motor activity.} 
\begin{figure}[h]
\vspace{0.7cm}

\includegraphics[width=0.49\textwidth]{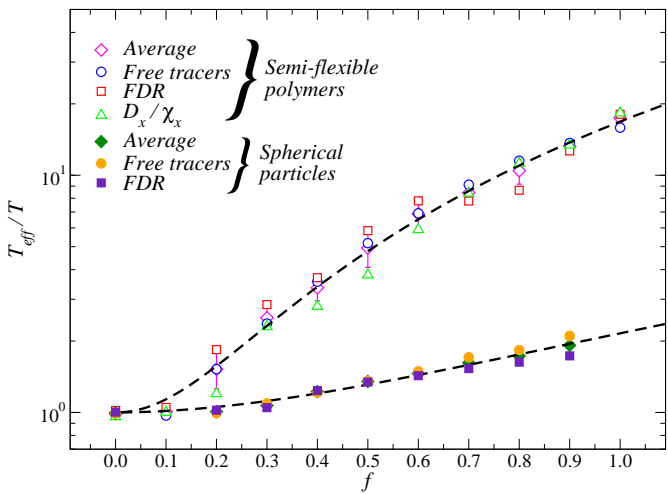}
\caption{ This figure summarizes the entire set of data that
we have generated, both for spherical self-propelled particles~\cite{loi08} and 
motorized semi-flexible polymers, on a linear-logarithmic scale. 
The different sets of data are indicated in the legend. Empirically, we 
find that all data are compatible with a power-law dependence of the effective 
temperature, $T_{eff}/T=1+\gamma f^2$ (dashed lines).
}
\label{fig:Teff-final} 
\end{figure}

In conclusion, in this section we have shown that $T_{eff}$ characterizes the 
chemical motor activity which we have schematized by the single parameter $f$. 
Figure~\ref{fig:Teff-final} collects all our results both for self-propelled 
particles~\cite{loi08} and motorized semi-flexible filaments. 
The data are presented in the form $T_{eff}/T$ against $f$ and we
also show (diamonds) average values with error bars calculated from the different
determinations of $T_{eff}$. 

The figure demonstrates that independent
determinations of $T_{eff}$ yield consistent results suggesting that, in these systems, 
there is a $T_{eff}$ parameter with  a thermodynamic meaning. The data for 
semi-flexible lines and spherical particles fall on different curves but there was
no reason for universality in this sense. They both tend to $T_{eff}=T$ in the 
limit of vanishing activation, as they should, and the deviation from the 
environmental temperature monotonically increases with increasing 
forcing. The data can be fitted by the empirical law 
$T_{eff}/T=1+\gamma f^2$  with $\gamma=15.41$ for filaments and $\gamma=1.18$ 
for particles, showed with dashed lines in the figure. We note that this finding 
seems to support our conjecture that the parameter $f$ plays here a role analogous 
to the Peclet number for colloidal active particles used in a few 
experiments~\cite{palacci10}, 
as discussed in Section~\ref{subsect:dynamics}.
\section{Conclusions and perspectives}
\label{sect:Conclusions}
The limits of the approach used in the present work are clear.  The
model we have developed is quite crude and does not take into account
explicitly any chemical activity nor particularly complex molecular
structure.  Moreover, we have chosen the most favorable conditions for
producing and observing the desired effects, at the risk of perhaps
being slightly unrealistic.  Nonetheless, we think that our results
are relevant for understanding very complex phenomena in real
systems. This conviction is based on two facts.

First, our work constitutes a genuine microscopic approach to active
matter.  In particular, no uncontrolled hypothesis has been made, the
physics of the passive system being under complete control: it
conforms to previously known results and it is only due to the effect
of conservative forces of very general nature. Also, the
implementation of the molecular motors, the exclusive source of non
equilibrium fluctuations in the system, is sufficiently general to be
representative of, at least, adamant motors. In conclusion, we expect
our results to be quite universal.  Second, as we discuss below, our
results are reminiscent of experimental findings reported in the
literature. From this point of view the approach presented here can
help both in rationalizing experimental observations and in suggesting
new research directions.

\subsection{Context}

A number of recent publications explored fluctuation dissipation
relations in biological systems or biologically motivated models. We
devote the following paragraphs to the discussion of results contained
in the papers that are relevant in the context of the present work.

Ziebert and Aranson~\cite{Ziebert} performed Langevin simulations of two-dimensional 
solutions of semi-flexible polar filaments interacting via molecular motors. 
They studied the rheology and structural properties in diluted samples and identified 
an ``active temperature'' $T_a$ via a phenomenological scaling of the elastic and loss 
moduli. $T_a$ increases with the average number of motors per filament and with 
the motor velocity. This definition is, though, not necessarily equivalent to the one that 
we used. Closer the our work are the following studies.

By comparing a hair bundle's spontaneous oscillations with its response to small mechanical 
stimuli, Martin {\it et al.}~\cite{Martin} demonstrated the breakdown of the FDT, and thus 
confirmed that a hair bundle's spontaneous movements are produced by energy-consuming 
elements within the hair cell. 

In~\cite{mizuno07} the fluctuation-dissipation relation in a model cell system formed by 
three fundamental components (myosin II, actin filaments, and cross-linkers) was studied 
in equilibrium and out-of-equilibrium conditions. FDT holds in the former and fails 
in the latter situation. In these papers, the violation of FDT was used 
as a proof of the non-equilibrium character of the dynamics induced by molecular 
motors. 

The effective temperature stemming from deviations from FDT has been
used to study the stability of motorized particles in
cytoplasm~\cite{shen04}. Interestingly enough, values of $T_{eff}$
that are lower than the ambient temperature were found when
considering motors that are aware of the free-energy landscape and act
only in the direction of lowering its instantaneous value. The motors
used in our study are not of this kind and the values of $T_{eff}$
obtained are consistently higher than the ambient $T$. Such an
``inversion'' also occurs in cases in which the initial configuration
is chosen to be to be one of equilibrium at a lower $T$ than the
working one~\cite{xy}.

A simple birth-death process and a general two-species interacting process, 
as simplified models of gene networks, were studied from the effective temperature 
perspective in~\cite{Wolynes}. The aim of this paper was to propose to use 
$T_{eff}$ as a means to distinguish between intrinsic and extrinsic noise sources.

Morozov {\it et al.}~\cite{Morozov} studied a model of the cytoskeletal network made 
of semi-flexible polymers subject to thermal and motor-induced fluctuations, 
modeled as either additive or multiplicative colored noise. 
They  found violations of the conventional FDT leading to a $T_{eff}$ that exceeds 
the thermodynamic temperature $T$ only in the low-frequency domain where motor 
agitation prevails over thermal fluctuations. This is consistent with our numerical 
findings. However, $T_{eff}$ in this phenomenological model is continuously frequency 
dependent while our data suggest that there is a low frequency regime 
in which it reaches a constant value, as in glassy systems.

Fluctuation-dissipation ratios were used to extract information about the degree of 
frustration, due to the existence of many metastable disordered states, in two self-assembly 
processes: the formation of viral capsids, and the crystallization of sticky discs~\cite{Jack}. 

Joly {\it et al}.~\cite{joly10} used numerical techniques to study the
non-equilibrium steady state dynamics of a heated crystalline
nanoparticle suspended in a fluid.  A two-temperature scenario holds
in this case. Indeed, by comparing the mobility to the velocity
correlation function, FDT approximately holds at short-time lags with
a temperature value that coincides with the kinetic one. In contrast,
at long-time lags data are compatible with the temperature estimated
by using the Einstein relation.

Ben-Isaac {\em et al.}~\cite{ben-isaac10} analyzed the Langevin dynamics of a single particle
randomly kicked by motors as a toy model of active matter. They focused on the deviations of 
$T_{eff}$ from $T$ and the distribution of velocities from Gaussian as parameters that
quantify the out of equilibrium nature of the dynamics, finding that while the former is robust 
the latter is not. Comparison to measurements in ATP-driven red-blood cells is also provided
in this paper.

Finally, let us sum up the results in the experimental work of Palacci
{\it et al}.~\cite{palacci10} that is, indeed, the closest to our
work.  These authors investigated the effective temperature by
following Perrin's analysis of the density profile in the steady state
of an active colloidal suspension under gravity. The active particles
used -- Janus particles -- are chemically powered colloids and the
suspension was studied with optical microscopy. The measurements show
that the active colloids are hotter than in the passive limit with an
effective temperature that increases as the square of the parameter
that controls activation, the Peclet number, a dependence that is
highly reminiscent of the $f^2$-dependence we found with our
simulations.

This summary proves that there is an increasing activity in the study of FDT violations 
and their interpretation in terms of effective temperatures in biophysics problems. 
A comprehensive experimental study of the thermodynamic properties, as we tried to present 
in this work within one system is, however, still lacking. In our opinion the active colloidal 
suspension used in~\cite{palacci10} appears as the most suitable experimental system where to 
deal with this kind of questions.

\subsection{Discussion}
Let us now critically review our findings.  We presented a study of
active matter made of semi-flexible filaments in interaction. In the
passive limit the conservative interaction parameters put the system
in a liquid phase. The equilibrium polymer melt has, though,
properties that are different from the ones found in the limits of
completely rigid or totally flexible polymers. For example, the
individual chains are not Gaussian.
 
Under the action of mild motor forces, the structure and dynamics are
significantly modified. The analysis of the chain radius of gyration
suggests that the polymers become more compact for increasing
forces. This is consistent with the behavior of thermodynamic
quantities that suggest crowding under increasing forces. On the other
hand, the study of the static structure factors suggests that the melt
gets more disordered --as in a passive sample in contact with hotter
baths-- and that chains become more folded when $f$ increases.  As a
consequence, the dynamics gets faster under stronger forces. This is
shown by the fact that the relaxation time, as extracted from the
evolution of the incoherent scattering function, decreases, while the
diffusion coefficient increases, for increasing $f$.

Next, we developed an in-depth study of the effective temperature of
the active system, by using a number of independent measurements. The
results of all these measurements in the semi-flexible polymer sample,
as well as the outcome of similar measurements in the self-propelled
point-particle model we analyzed in~\cite{loi08}, are summarized in
Fig.~\ref{fig:Teff-final}.

The dependence of $T_{\sc eff}$ on $f$ conforms to our intuition.  We
dealt with motors that act randomly -- and ignore their effect on the
energy landscape of the system. They have the effect of disordering
the sample and, consequently, the stronger $f$ the higher $T_{eff}$.
In their study of point-particle active matter~\cite{shen04}, Shen and
Wolynes argued that motors that act selectively, only when they
contribute to decreasing the system's free-energy density, lead to out
of equilibrium dynamics with lower effective temperature than the
environmental one.  The interpretation is that this kind of motor
orders the sample more than the passive system would be in equilibrium
at the working bath temperature.  A similar inversion has been
observed in relaxing passive systems driven from an ordered initial
equilibrium condition to a point in parameter space in which the
evolution slowly tends to reach a less ordered final equilibrium
state~\cite{xy}.  In this case, the dynamic configurations are also
more ordered than the target equilibrium ones and, as a consequence,
their effective temperature is lower than the one of the bath.

The results discussed in this paper and in Ref.~\cite{loi08} give a
different perspective on the dynamics of active matter than the
coarse-grained hydrodynamic
descriptions~\cite{marchetti,Hatwalne04,Kruse}.  While in the
mode-coupling study in~\cite{Hatwalne04} a strong frequency dependence
of the effective temperature was found, in our simulations the
behavior observed is, in contrast, very similar to the one advocated
in~\cite{cugliandolo97} to be the one of interacting glassy-like
systems evolving in a small entropy production limit. Indeed, the
global behavior of $T_{\sc eff}$ in the active matter polymer system
resembles the one computed in a variety of atomic models in the glassy
regime~\cite{glassy-Teff}, sheared super-cooled
liquids~\cite{Berthier-Barrat}, driven vortex systems~\cite{Kolton} or
even vibrated granular matter~\cite{Makse}. Independent measurements
of $T_{\sc eff}$ yield consistent results in the atomic and molecular
cases, as shown in Fig.~\ref{fig:Teff-final}, giving support to the
idea that $T_{\sc eff}$ can be considered to be a thermodynamic
object, at least in the weakly driven cases studied here ($f\leq
1$). It would be interesting to test whether a similar phenomenology
arises in lattice models of
Vicsek-type~\cite{Vicsek,Chate} with various independent measurements of 
$T_{eff}$ yielding the same result. On the experimental front,
it would be welcome to extend the analysis of $T_{eff}$ in the dilute
active suspension of artificial swimmers studied in~\cite{palacci10}
or search for $T_{eff}$ in artificial samples such as the ones
recently designed for their tunability~\cite{Dauchot}.

Finally, it would be interesting to explore the possibilities offered by the approach
described in this work to clarify the extremely complex phenomenology
of cell mechanical stability~\cite{cyto}. In particular, one could try to give an 
answer to the question as to how  mechanical properties like 
elastic moduli change with motor activity and whether these changes can also be rationalized 
in terms of the effective temperature.

\vspace{0.5cm}
\acknowledgments
We thank J.-L. Barrat, G. Gonnella and A. Parmeggiani for useful comments. 
This research was supported in part by the National Science Foundation under 
Grant No. NSF PHY05-51164.

\begin{thebibliography}{99}
\bibitem{footnote1}
The persistence length, $\lambda_P$, is usually defined as the characteristic length scale 
for the exponential decay of the two-point correlation function between unit tangent vectors 
along a polymer and quantifies its degree of stiffness. Only pieces of the polymer of length 
shorter than persistence length appear rigid and maintain an average straight conformation 
with a preferred tangent direction. In contrast, pieces of size larger than $\lambda_P$ 
appear completely flexible. 
The contour length, $\lambda_C$, is defined as the maximum end-to-end distance of a linear 
polymer chain. In the present semi-flexible case the persistence length is comparable to the 
contour length.
\bibitem{momenta-note}
We recall that variance, skewness and kurtosis characterize width, asymmetry 
and peakedness of a probability distribution, respectively.
\bibitem{note}
Note that this is not the same as defining the effective temperature as the long time limit of
$\chi(t)/C(0)$. As noted in Ref.~\cite{Berthier-Barrat}, the latter method corresponds to determine
$T_{eff}$ as the average slope of the parametric plot, therefore neglecting the separation of time 
scales and finding inaccurate results~\cite{ono02}.
\bibitem{note-referee}
We are indebted to an anonymous referee of~\cite{loi11}
who raised our attention on this point.
\end{thebibliography}
\end{document}